\documentclass[a4paper, amsfonts, amssymb, amsmath, reprint, showkeys, nofootinbib]{revtex4-1}
\usepackage[english]{babel}
\usepackage[utf8]{inputenc}
\usepackage[colorinlistoftodos, color=green!40, prependcaption]{todonotes}
\usepackage[pdftex, pdftitle={Article}, pdfauthor={Author}]{hyperref}% For hyperlinks in the PDF
\usepackage{float}
\usepackage{graphicx}
\usepackage{yhmath}
\usepackage{subfigure}
\usepackage{mathdots}
\usepackage{MnSymbol}

\begin{document}
\title{Physics-Enhanced Bifurcation Optimisers: All You Need Is a Canonical Complex Network }
	\author{Marvin Syed and Natalia G. Berloff}
	\email[correspondence address: ]{N.G.Berloff@damtp.cam.ac.uk}	
	\affiliation{Department of Applied Mathematics and Theoretical Physics, University of Cambridge, Cambridge CB3 0WA, United Kingdom}
	
	\begin{abstract}{Many physical systems with the dynamical evolution that at its steady state gives a solution to optimization problems were proposed and realized as promising alternatives to conventional computing. Systems of oscillators such as  coherent Ising and XY machines based on lasers, optical parametric oscillators,  memristors,  polariton and photon condensates are particularly promising due to their scalability, low power consumption and room temperature operation. They achieve a solution via the bifurcation of the fundamental supermode that globally minimizes either the power dissipation of the system or the system Hamiltonian. We show that the canonical  Andronov-Hopf networks can capture the bifurcation behaviour of the physical optimizer. Furthermore, a continuous change of variables transforms any physical optimizer into the canonical network so that the success of the physical XY-Ising machine depends primarily on how well the parameters of the networks can be controlled. Our work, therefore, places different physical optimizers in the same mathematical framework that allows for the hybridization of ideas across disparate physical platforms.  }
	\end{abstract}
	
	\maketitle
	{\it Introduction.}
    Optimization problems are ubiquitous in technology and applications, from machine learning and artificial intelligence to industrial designs of vehicles, new materials, and drugs \cite{kitagawa2004applications, odili2017applications, paschos2014applications}. The complexity of such problems grows fast, often with an exponential increase in the number of candidate solutions with the number of unknowns. Given the importance of finding a reasonable solution quickly while searching a large hyperspace of an ever-increasing number of variables,  there have been intense efforts to design analogue physical hardware capable of performing this task. The feasibility of this task has been assured by the complexity theory stating that most if not all  optimization   problems can be mapped into universal spin Hamiltonians with a polynomial overhead on the number of additional variables (spins) \cite{Barahona1982, Cubitt_universality,lucas2014ising}.

    Suppose a physical system can be arranged and controlled, so   Hamiltonian is globally minimised. In that case, this physical system can potentially be used as an unconventional physics-enhanced computing device for solving these types of tasks. XY, $H_{XY}$, Ising, $H_I$ and k-local, $H_{K}$ classical Hamiltonians formulated as 
\begin{eqnarray}
    H_{XY}&=&\min_{\{\theta_i\in[0,2 \pi)\}} -\sum_i\sum_j J_{ij} \cos(\theta_i-\theta_j),\label{xy}\\
     H_{I}&=&\min_{\{s_i=\pm 1\}} -\sum_i\sum_j J_{ij} s_i\cdot s_j, \label{i}\\
     H_K &=&\min_{\{s_i=\pm 1\}} -\sum_{i,j,...,k} J_{ij,...,k} s_i\cdot s_j\cdot\cdot\cdot s_k,
    \label{k}
\end{eqnarray}
are all universal, meaning also that for the most general coupling matrix $J,$ finding the global minimum of the corresponding Hamiltonian is an NP-hard problem that requires exponentially fast growing resources. 

 Various physics-based Ising and XY spin minimisers -- XY-Ising machines --  have been created using  lasers \cite{babaeian2019single, pal2020rapid, parto2020realizing}, optical parametric oscillators \cite{yamamoto2017coherent, inagaki2016coherent, mcmahon2016fully}, superconducting qubits \cite{johnson2011quantum, denchev2016computational, arute2019quantum}, memristors \cite{cai2020power}, trapped ions \cite{kim2010quantum}, polariton condensates \cite{berloff2017realizing, kalinin2020polaritonic}, and photon condensates \cite{vretenar2021controllable} among many others.
 Some of these solvers achieve minimisation using the underlying physical principle of minimum power dissipation subject to constraints on voltage, amplitude, gain, etc. \cite{vadlamani2020physics}. Others use quantum or classical annealing. For instance, in the vicinity of a power-dissipation minimum, the system evolves in time following the gradient of the power-dissipation function. However, to offer a computational advantage, the successful XY-Ising machine cannot be based on gradient descent alone. The success of the XY-Ising devices depends on their ability to search the low-energy part of the spin Hamiltonian without being trapped by the local minima. This is through the {\bf bifurcation} of an additional degree of freedom --  the amplitude of the laser field or condensate wavefunction -- that the system selects the fundamental mode, and, therefore, the minimum of the power-dissipation function.  
 
 The amplitude bifurcation often is a key to other physical minimisation principles, for instance, when the system 
 evolves adiabatically or anneals to the minimum of the system Hamiltonian. Annealing here concerns the changing Hamiltonian during the system evolution that can occur adiabatically \cite{hauke2020perspectives} or not \cite{kamaletdinov2021quantized}. Toshiba bifurcation machine is an example of such bifurcation on the route to minimise the Ising Hamiltonian \cite{goto2016bifurcation,tatsumura2019fpga,goto2021high}.
 
 Many lasers, photonic, polaritonic  and biological systems exhibit the so-called Andronov-Hopf bifurcation at the threshold that leads to the birth of the limit cycle out of an equilibrium point. The canonical model describing this bifurcation is the network of the Andronov-Hopf oscillators (AHO) that  can be written   in the most general form  as   
 \begin{equation}
    \dot{\psi_i}= (\gamma_i + i \omega_i)\psi_i - (\sigma_i +i U_i)|\psi_i|^2 \psi_i + \sum_{j\ne i} Q_{ij}(\psi_j).
    \label{cc}
\end{equation}
Here $\psi_i(t)$ is a complex function of time that characterises the state of the $i-$th oscillator,  $Q_{ij}$ describes the coupling between the $i-$ and $j-$ the oscillators and $\gamma_i, \omega_i, \sigma_i, U_i$ represent the effective gain, self-frequency, nonlinear dissipation and self-interactions, respectively. In particular, polaritonic networks or laser \cite{pal2020rapid} show a potential of controlling all these terms independently \cite{kalinin2019polaritonic, kalinin2020polaritonic}. In what follows, we show that Eq.~(\ref{cc}) is a general framework that describes different classes of physical optimizers considered in the literature and formulate when this model achieves the minimum of universal spin Hamiltonian.    

\section{Andronov-Hopf oscillators as coherent  XY-Ising machines}
The  Hopfield models are perhaps the best-known networks used to minimise the Ising Hamiltonians. They are also used to describe the dynamics of the coherent Ising machines (CIMs) \cite{wang2013coherent, inagaki2016coherent}. They are trivially reduced to the AHO networks under parametric pumping that projects the phases to $0$ or $\pi$. In CIMs, the state $x_i$ corresponds to the in-phase amplitude of the $i-$th oscillator pulse, whose dynamics are described by 
\begin{equation}
    \frac{d x_i}{d t} = (p-1 - x_i^2) x_i + \xi\sum_{j\ne i}G_{ij} x_j,
    \label{CIM}
\end{equation}
where $p$ is the photon injection rate and $\xi$ is the suitable scaling factor. Equation (\ref{CIM}) coincides with Eq.~(\ref{cc}) if $\theta_i \in \{0,\pi\}.$ In this case,  $\psi_i=r_i \exp[i \theta_i + i\omega_i].$   To see this we let $x_i=r_i \exp[i \theta_i],\theta_i=\{0,\pi\},  \omega_i=\omega, \gamma_i=p-1, \sigma_i=1, U=0, Q_{ij}(\psi_j)=\xi G_{ij} \psi_j$. The projection of phases onto $0$ or $\pi$ will be automatically  achieved if only the real parts of the fields are coupled so the coupling in Eq.~(\ref{cc}) takes the form $Q_{ij}(\psi_j)=\xi G_{ij} (\psi_j + \psi_j^*).$

Various modifications of the CIM and/or the Hopfield networks can be accommodated by Eq.~(\ref{cc}).  For instance, the success of the  Hopfield network minimisers (coherent Ising machines)   is improved with the introduction  of the chaotic amplitude method \cite{leleu2020chaotic} that anneals the coupling terms as $\xi=\epsilon_i$ where $\epsilon_i$ depends on how far away each oscillator is from its saturation amplitude. These annealing schedules can be  introduced into the canonical form of Eq.~(\ref{cc}) and, therefore, in principle, realised by  any optical  network described by Eq.~(\ref{cc}) that offers that kind of control.

The minimisers of the higher order binary optimization problems can be obtained by the higher order Hopfield networks \cite{stroev2021discrete}   if the coupling term in Eq.~(\ref{cc}) is replaced by the higher order coupling leading to 
\begin{equation}
\begin{split}
    \dot{\psi_i} = (\gamma_i + i \omega_i)\psi_i - (\sigma_i + i U_i)|\psi_i|^2 \psi_i\ + \\ \sum_{j,k,...l} J_{ijk...l} \psi_j\psi_k...\psi_l^*.
    \label{cc-higher}
\end{split}
\end{equation}
No projection of the phases to the discrete values $0$ and $\pi$ are needed in this case, as such projection is automatically achieved by mixing $\psi_j$ and $\psi_k^*$ in the coupling terms as was argued in \cite{stroev2021discrete}.

 The development of the XY-Ising machines postdated extensive research on networks of  neural oscillators near multiple Andronov-Hopf bifurcation points. In particular, weakly interacting networks were proposed as oscillatory neurocomputers capable of emulating an associative memory network. Networks consist of $N$ neural oscillators comprised of two populations of neurons excitatory, described by a scalar function of time $x_i(t)$ and inhibitory, $y_i(t)$, that evolve according to the dynamical equations \cite{hoppensteadt1996synaptic,hoppensteadt1996synapticB}
 \begin{eqnarray}
 \dot{x_i}&=&f(x_i, y_i, \lambda_i) +  \epsilon p_i(x_1, y_1,\cdot\cdot\cdot, x_n,y_n, \epsilon), \label{xi}\\
 \dot{y_i}&=&g(x_i, y_i, \lambda_i) + \epsilon q_i(x_1, y_1,\cdot\cdot\cdot, x_n,y_n, \epsilon), \label{yi}
 \end{eqnarray}
 where $\lambda_i$ is a bifurcation parameter and  $\epsilon$ is a small parameter describing the strength of interactions between neurons. Functions $f, g, p_i,$ and  $ q_i$ describe the self-evolution and couplings among the neurons. 

 The dynamical system described by Eqs.~(\ref{xi},\ref{yi}) with $\epsilon=0$ is near an Andronov-Hopf bifurcation if the Jacobian matrix
\begin{equation}
  \frac{D(f,g)}{D(x,y)}= \biggl(\begin{matrix}\frac{\partial f}{\partial x},\frac{\partial f}{\partial y}\\
    \frac{\partial g}{\partial x},\frac{\partial g}{\partial y}\end{matrix}\biggr)
\end{equation}
has a  pair of purely imaginary eigenvalues that we denote $\pm i \Omega.$ The corresponding column (row) eigenvectors we  denote as ${\bf v}$ and ${\bf \bar{v}}$ (${\bf w}$ and ${\bf \bar{w}}$) and form a matrix $V=({\bf v},{\bf {\bar v}})$. Changing the variables to
\begin{equation}
  \biggl(\begin{matrix} x_i(t)\\ y_i(t) \end{matrix}\biggr) =\sqrt{\epsilon} V \biggl(\begin{matrix} \exp[i \Omega t] \psi_i(\tau)\\ \exp[-i \Omega t] \psi_i^*(\tau) \end{matrix}\biggr) + {\cal O}(\epsilon),
  \label{transform}
\end{equation}
introducing slow time $t=\epsilon \tau$ and considering the dynamics near 
 multiple Andronov-Hopf bifurcation points with $\lambda \rightarrow  \epsilon \lambda$ reduces 
 Eqs.~(\ref{xi},\ref{yi}) to the canonical form of Eq.~(\ref{cc}) in the order $O(\sqrt{\epsilon})$ \cite{hoppensteadt1997weakly}. 
The parameters $\gamma_i, \sigma_i, U_i$ and $\omega_i$ in Eq.~(\ref{cc}) depend on $\lambda$ and $\Omega$ (so on the structure of $f$ and $g$) and the  couplings become  $Q_{ij}(\psi_j)={\bf w} \cdot D (p_i,q_i)/D(x_j,y_j) \cdot {\bf v} \psi_j$.

There are two further popular examples of XY-Ising machines that operate  near 
 multiple Andronov- Hopf bifurcation points.

Coupled microelectromechanical systems (MEMs)  \cite{hoppensteadt2001synchronization} are governed by 
\begin{equation}
    \ddot{x_i} + F(x_i,\lambda_i)\dot{x_i} + G(x_i)=\sum_{j\ne i} (e_{ij} \dot{x_j} + k_{ij} x_j),
    \label{mems}
\end{equation}
where $x_i$ is the displacement from the rest position, $F$ and $G$ are damping and stiffness parameters,  $e_{ij}$ [$k_{ij}$] are electric conductances [mechanical spring] constants coupling   the $i$-th and the $j$-th oscillators. Clearly, Eq.~(\ref{mems}) can be written as  Eqs.~(\ref{xi},\ref{yi}) by letting $y_i=\dot{x_i}.$ The reduction to the canonical AHO is achieved by writing $\psi_i=\dot{x_i} + i \sqrt{\lambda} x_i.$ \cite{hoppensteadt2001synchronization}. The relationship between the coefficients of Eq.~(\ref{mems}) and the coefficients of AHO in Eq.~(\ref{cc}) are given in \cite{hoppensteadt2001synchronization}.

 The dynamics of Eq.~(\ref{mems}) can be viewed as a particular case of  gradient descent accelerated by momentum, also known as Nesterov’s accelerated gradient method \cite{nesterov2003introductory,su2014differential}:
\begin{equation}
    \ddot{x_i} + \lambda(t) \dot{x_i} + \xi \nabla E({\bf x}) = 0,
    \label{nesterov}
\end{equation}
where $\lambda(t)>0$ is the friction coefficient and $E({\bf x})=-\frac{1}{2}\sum_{i,j}J_{ij} x_i x_j$. $E({\bf x})$ becomes the Ising Hamiltonian if we replace $x_i$ with ${\rm sign}(x_i).$ 
More generally, Eq.~(\ref{nesterov}) is a special case of a conformal Hamiltonian system of the form
\begin{eqnarray}
    \dot{{\bf x}}&=&\nabla_{\bf y} H({\bf x},{\bf y}),\label{conformal1}\\
    \dot{{\bf y}}&=&-\nabla_{\bf x} H({\bf x},{\bf y})-\lambda {\bf y}.
    \label{conformal2}
\end{eqnarray}
If $H$ is separable Hamiltonian,  $H({\bf x},{\bf}y)=\frac{1}{2} \lambda ||{\bf y}||^2_2 +\xi E({\bf x}),$ we get back to Eq.~(\ref{nesterov}) \cite{celledoni2021structure}.
The reduction to AHO is obtained similar to \cite{hoppensteadt2001synchronization} using $\psi_i=\dot{x_i} + i \sqrt{\lambda} x_i$ giving $Q_{ij}=i\frac{1}{2}\xi J_{ij} \psi_j, \omega_i=\sqrt{\lambda}, \gamma_i=-\frac{1}{2}(\lambda-\lambda_H), $ where $\lambda_H$ is the threshold for the bifurcation. The qubic terms $\sigma_i$ and $U_i$ do not appear after the transformation at this order, but their introduction into the equations helps to saturate the gain faster. 

Toshiba bifurcation machine \cite{goto2016bifurcation,tatsumura2019fpga,goto2021high} has demonstrated an improvement over the CIM by employing  adiabatic evolutions of energy conservative systems motivated by  purely adiabatic quantum annealing. Its dynamics is governed by
\begin{eqnarray}
    \dot{x_i}&=&a_0 y_i, \label{toshiba1}\\
    \dot{y_i}&=&-(a_0-a(t)) x_i + \xi \sum_{j\ne i} J_{ij} x_j,\label{toshiba2}
\end{eqnarray}
where the state of each oscillator is described by two real variables $x_i$ and $y_i $  and the annealing is performed by letting $a=a(t)$ approach $a_0$ as $t\rightarrow t_{\max}$, while $t_{\max}$ is the terminal time of the dynamics. As in CIM, the spins are associated with the $s_i={\rm sign}(x_i)$ at the fixed point of the dynamics. 

To get the canonical AHO equations at the onset of bifurcation we assume that $a_0-a$ is constant and write
\begin{equation}
    x_i=\frac{1}{2}(\psi_i + \psi_i^*)\sqrt{a_0}, \quad y_i=\frac{1}{2 i} (\psi_i-\psi_i^*) \sqrt{a_0-a},
\end{equation}
so that $\psi_i=x_i/\sqrt{a_0} +i y_i/\sqrt{a_0-a}.$ The AHO network of  Eq.~(\ref{cc}) becomes
\begin{equation}
    \dot{\psi_i}=-i\sqrt{a_0}\sqrt{a_0-a} \psi_i + \frac{i \xi}{\sqrt{a_0-a}}\sum_{j\ne i} J_ij (\psi_j + \psi_j^*).
    \label{tbcc}
\end{equation}
The canonical form given by Eq.~(\ref{tbcc}) (Eq.~(\ref{cc})) is capable, therefore, of capturing the dynamics of the system close to the bifurcation point.

\begin{figure*}[!t]
    \centering
    \includegraphics[width=\textwidth]{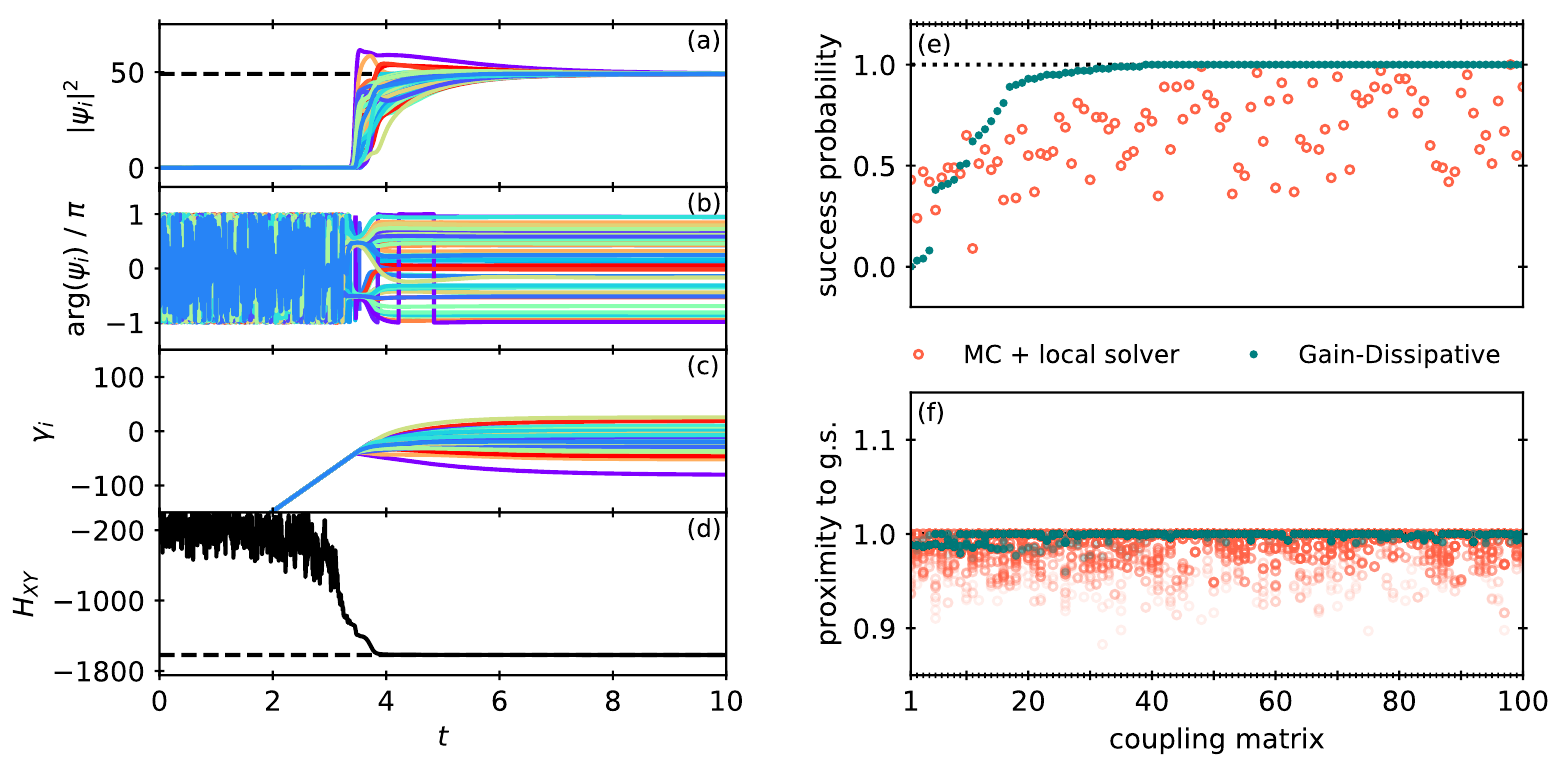}
    \caption{XY minimization using Eqs.\,\eqref{cc} and \eqref{gain} with additive noise on random $50\times50$ coupling matrices with elements drawn from the uniform distribution on $[-10,10]$. (a)-(d) The time evolution of one specific instance is shown to illustrate the typical behaviour of the system. (e) The success probabilities for minimising the XY Hamiltonian on 100 different random coupling matrices (as defined above) are shown. A success is here defined as coming within a margin of 99\% of the minimal energy $\min H_{XY}$ found in all of the runs. Teal dots represent the success probabilities obtained by solving Eqs.\,\eqref{cc} and \eqref{gain}, while orange circles represent the ones obtained by using a quasi-Newton method. Each method was initialized with 100 different random initial conditions (and noise seeds for the AHO based method) to obtain statistics. In (f) we show the proximity to the ground state for all runs, where the proximity is defined as the ratio of the found minimum and $\min H_{XY}$.}
    \label{fig1}
\end{figure*}
In the work of any physical XY-Ising machine, the dynamics near the Andronov-Hopf bifurcation are the most important for optimization as well as for the associative memory task because each oscillator must be near a bifurcation to make a nontrivial contribution to the entire network dynamics as follows from the fundamental theorem of weakly connected neural network theory \cite{hoppensteadt1997weakly}. The difference in the demonstrated behaviour of various XY-Ising machines comes, therefore, not from the key mathematical properties of the operation of such devices but the annealing schedule of the parameters.

Next, we will clarify the relationship between Eq.(\ref{cc}) and the XY Hamiltonian minimisation (while the correspondence with other classical spin Hamiltonians follows when one takes into account the structure of the spins and the coupling terms). 
Let the coupling term $Q_{ij}(\psi_j)=s_{ij} \exp[i \phi_{ij}] \psi_j$ and $\psi_i=r_i \exp[i \theta_i]$ so that Eq.~(\ref{cc}) can be written in polar coordinates
\begin{eqnarray}
 \dot{r_i}&=&\gamma_i r_i - \sigma r_i^3 + \sum_{j\ne i} s_{ij} r_j \cos(\theta_i-\theta_j-\phi_{ij}),\label{r}\\
 \dot{\theta_i}&=&\omega - U r_i^2 -\frac{1}{r_i} \sum_{j\ne i} s_{ij} r_j \sin(\theta_i-\theta_j-\phi_{ij}),\label{theta}
\end{eqnarray}
where we assumed that $\omega_i=\omega, \sigma_i=\sigma, U_i=U.$
The relationship with the XY  models can be obtained by either assuming that $\sigma \gg \max s_{ij}$ \cite{hoppensteadt2001synchronization} or by using a feedback on the gain coefficients \cite{kalinin2018gain, kalinin2018networks,kalinin2018global}. We will discuss both approaches below.

If $\sigma \gg \max s_{ij}$ and all oscillators are pumped with the same intensity $\gamma_i=\gamma$, the last term on the right-hand side of Eq.~(\ref{r}) is negligible in comparison with other terms, so that the oscillators amplitudes take on a stationary values $r_i=r=\sqrt{\gamma/\sigma}.$ Equation (\ref{theta}) reduces to the Kuramoto-Sakaguchi model of  oscillators with identical natural frequency $\tilde{\omega}=\omega -U \gamma/\sigma$ and natural phase lag $\phi_{ij}$
\begin{equation}
    \dot{\theta_i}=\tilde{\omega} - \sum_{j\ne i} s_{ij} \sin(\theta_i-\theta_j-\phi_{ij}).\label{kuramoto-S}
\end{equation}
If the couplings are real, so that $\phi_{ij}\in \{0,\pi\}$ then Eq.~(\ref{kuramoto-S}) reduces to the Kuramoto model 
\begin{equation}
    \dot{\theta_i}=\tilde{\omega} - \sum_{j\ne i} J_{ij} \sin(\theta_i-\theta_j), \label{kuramoto-ising}
\end{equation}
where $J_{ij}=\pm s_{ij}$. Starting from any initial condition Eq.~(\ref{kuramoto-ising}) follows the gradient descent to the minimum of the classical XY Hamiltonian $H_{XY}$.
It was shown that the network of oscillators reproducing Eqs.~(\ref{cc})  and using the Hebbian learning rule  has associative memory similar to that of Hopfield–Grossberg networks, but a greater memory capacity \cite{izhikevich2000computing}. While the gradient descent is sufficient for pattern recognition and other associative memory applications, optimization tasks require the system to be able to escape the local minima in its search for the global one. 
This search benefits from unequal and dynamically changing amplitudes that bifurcate from zero as the gain increases. However, these amplitudes must all reach the same value at the steady state to minimise the  Hamiltonian with the given coupling matrix \cite{kalinin2018networks}. This is achieved by complementing Eqs.~(\ref{cc})  with time-evolving gains
\begin{equation}
    \dot{\gamma_i}=\tilde{\epsilon} (1 - r_i^2),
    \label{gain}
\end{equation}
where the parameter $\tilde{\epsilon}$ controls the rate of change of $\gamma_i$ with respect to the amplitudes of the oscillators. 

As the network of oscillators approaches the steady state, the amplitudes approach one, while the phases start evolving according to Eqs.~(\ref{kuramoto-S},\ref{kuramoto-ising})  and the total occupancy (mass) of the system of $N$ oscillators becomes
\begin{equation}
    N\sigma=\sum_{i=1}^N \gamma_i +H_{XY}.
\end{equation}
It follows that if the total effective gain $\sum_{i=1}^N \gamma_i$ is globally minimised, then $H_{XY}$ is also globally minimized.

\section{XY-Ising machines for global minimisation}
In the previous section, we argued that the physical optimisers could be reduced to the canonical complex AHO networks in the vicinity of the bifurcation. However, all considered optimisers involve time-varying (annealed) parameters, so they all will have different dynamics before and after the bifurcation. We argue, however, that as follows from the fundamental theorem of weakly connected neural network theory \cite{hoppensteadt1997weakly} only the region close to bifurcation is essential for the global minimisation; therefore, we can always choose the annealing schedule to bring different systems to the same behaviour at the bifurcation point, and, therefore, to the same solution. In this section, we illustrate this by using numerical simulations of the canonical complex  AHO networks for XY and Ising Hamiltonian minimisation and demonstrate that AHO behaviour corresponds to the operation of vastly different machines considered in the previous section if annealing schedules are suitably chosen.  

{\it XY machine.} For XY minimization we use Eqs.~(\ref{cc}) with additive noise and (\ref{gain}) with $\omega_i=0, \sigma_i=1, U_i=0$ and $Q_{ij}(\psi_j)=J_{ij} \psi_j. $ Figs.\,1(a-d) illustrate the typical numerical evolution of the system. Figs.\,1(e-f) show the statistics of finding the global minimum compared to a brute force Monte Carlo method.
In most cases, the AHO finds the global minimum with a very high probability. In contrast, the system still seeks out a local minimum close to the ground state for the coupling matrices where the success probability of finding the true ground state is low.
Comparison to a quasi-Newton method, on the other hand, shows that the actual distribution of local minima is far more spread out.

{\it Comparison of Ising machines.}
 To illustrate that AHO captures the behaviours of the Coherent Ising Machines and the Toshiba Bifurcation Machines, we numerically simulate Eq.~(\ref{cc}) using the mapping presented   and compare the results with the dynamical behaviour of Eqs.~(\ref{CIM}) and Eqs.~(\ref{toshiba2}) on two different graphs. 
\begin{figure*}[!t]
    \centering
    \includegraphics{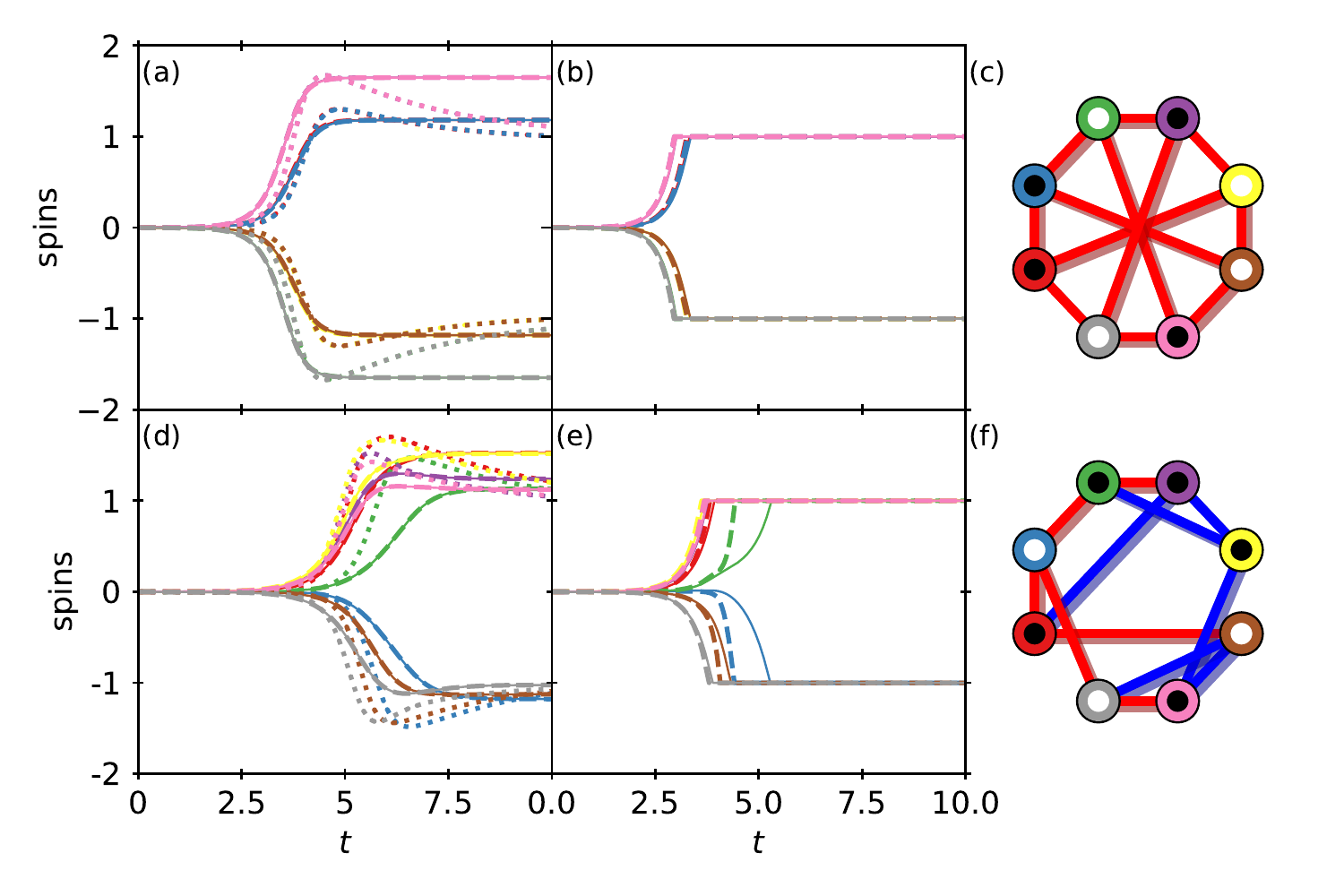}
    \caption{Comparison of the dynamical behaviour of Eqs.\,\eqref{cc}, \eqref{CIM}, and \eqref{toshiba1} and \eqref{toshiba2} when solving the Ising model on a Moebius ladder graph with solely negative couplings (a-c) and on a random 3-regular graph with mixed negative and positive couplings (d-f). In (a,b,d,e), each color represents one of the eight spins. The respective graphs used are shown in (c,f) with blue (red) edges signifying a coupling of $J_{ij} = +1$ ($J_{ij} = -1$) and black and white vertices representing positive and negative spin orientations respectively. All equations were integrated using the 4th order Runge-Kutte method with a step size of h=0.01 and 1000 time steps. Every method starts with identical initial condtions, where each spin is drawn from the uniform distribution on $[-x_0, x_0]$ and $x_0 = 5 \times 10^{-4}$. Thin solid curves in (a,d): CIM described by Eq.\,\eqref{CIM} with annealing schedule $p(t) = \min\,\{ t, 1\}$ and scaling factor $\xi=1$. Dashed in curves in (a,d): Canonical AHO described by Eq.\,\eqref{cc} with $Q_{ij}(\psi_j) = J_{ij} (\psi_j + \psi_j^*)$, $\gamma_i(t) = p(t) - 1$, $\sigma_i=1$, and $U_i=\omega_i=0$. Dotted curves in (a,d): Identical to the AHO before but $\gamma(t)$ evolves according to Eq.\,\eqref{gain} with $\tilde{\epsilon} = 0.4$ and $\gamma_i(t=0) = -1$. Thin solid curves in (b,e): Toshiba bifurcation machine described by Eqs.\,\eqref{toshiba1} and \eqref{toshiba2} with $a_0 = 1$, $a(t) = \min\,\{t,1\}$, and $\xi = 3.5$. Dashed curves in (b,e): Canonical AHO with $\omega_i(t) = -\sqrt{a_0} \sqrt{a_0 -a(t/5)}$, $Q_{ij}(\psi_j) = i\xi(t) J_{ij} (\psi_j + \psi_j^*)$, where $\xi(t) = 2\sqrt{a_0} / (\sqrt{a_0 - a(t/4)} + 0.05)$, and $U_i = \sigma_i = \gamma_i = 0$. Note that for both methods used in (b,e) we implement perfectly inelastic walls as in \cite{goto2021high}.}
    \label{fig:aho_comp}
\end{figure*}
The results are displayed in Fig.\,\ref{fig:aho_comp}, where we compare the time evolution of the CIM and Toshiba Bifurcation machine described by Eqs.\,(\ref{CIM}) and \eqref{toshiba1}-\eqref{toshiba2} respectively with that of the canonical AHO described by Eq.\,(\ref{cc}).
For both linearly annealed gains, $\gamma_i$ and gains controlled by Eq.\,\eqref{gain}, the canonical AHO manages to replicate the behavior of the CIM and Toshiba bifurcation machine close to the bifurcation.
When minimising the Ising energy on the random graph used in Fig.\,\ref{fig:aho_comp}(d-e), we observe that some of the spins exhibit a delayed bifurcation due to frustration effects. The evolution of AHO captures a delayed bifurcation as well. 

\section{Conclusions}
High parallelism, processing speed, shared memory, energy efficiency and other advantages of analogue physical simulators led to the development of a plethora of competing platforms and physics-inspired optimisation methods. The analogue mode of operation of such platforms typically emulates interacting dynamical systems and their behaviour near critical regimes, such as bifurcations, determine their optimisation properties.  From the mathematics of dynamical systems  we know that many diverse systems behave similarly close to the bifurcation points and, therefore, share similar universal description  by canonical models. Such canonical models are capable of describing the systems' operation near criticality even when the exact mathematical description of that system is not known or too complex. Here, we show how the popular physical platforms used as optimisers can all be described as canonical AHO networks. 

When the physical platforms are presented by vastly different mathematical formulations, it is hard to directly compare the existing methods and the performance of such platforms. Such comparison requires optimal parameters for each platform that can be different for different problem structures. However, as we argue in our paper, as long as the primary mechanism for optimisation is based on the behaviour at the bifurcation point, the canonical complex AHO networks can represent all such models. The performance of the method and the physical platform depends only on the annealing schedule of the coefficients and the feasibility to realise such controls in practice.

\bibliography{references}{}

\providecommand{\noopsort}[1]{}\providecommand{\singleletter}[1]{#1}%
\begin{thebibliography}{41}%
\makeatletter
\providecommand \@ifxundefined [1]{%
 \@ifx{#1\undefined}
}%
\providecommand \@ifnum [1]{%
 \ifnum #1\expandafter \@firstoftwo
 \else \expandafter \@secondoftwo
 \fi
}%
\providecommand \@ifx [1]{%
 \ifx #1\expandafter \@firstoftwo
 \else \expandafter \@secondoftwo
 \fi
}%
\providecommand \natexlab [1]{#1}%
\providecommand \enquote  [1]{``#1''}%
\providecommand \bibnamefont  [1]{#1}%
\providecommand \bibfnamefont [1]{#1}%
\providecommand \citenamefont [1]{#1}%
\providecommand \href@noop [0]{\@secondoftwo}%
\providecommand \href [0]{\begingroup \@sanitize@url \@href}%
\providecommand \@href[1]{\@@startlink{#1}\@@href}%
\providecommand \@@href[1]{\endgroup#1\@@endlink}%
\providecommand \@sanitize@url [0]{\catcode `\\12\catcode `\$12\catcode
  `\&12\catcode `\#12\catcode `\^12\catcode `\_12\catcode `\%12\relax}%
\providecommand \@@startlink[1]{}%
\providecommand \@@endlink[0]{}%
\providecommand \url  [0]{\begingroup\@sanitize@url \@url }%
\providecommand \@url [1]{\endgroup\@href {#1}{\urlprefix }}%
\providecommand \urlprefix  [0]{URL }%
\providecommand \Eprint [0]{\href }%
\providecommand \doibase [0]{http://dx.doi.org/}%
\providecommand \selectlanguage [0]{\@gobble}%
\providecommand \bibinfo  [0]{\@secondoftwo}%
\providecommand \bibfield  [0]{\@secondoftwo}%
\providecommand \translation [1]{[#1]}%
\providecommand \BibitemOpen [0]{}%
\providecommand \bibitemStop [0]{}%
\providecommand \bibitemNoStop [0]{.\EOS\space}%
\providecommand \EOS [0]{\spacefactor3000\relax}%
\providecommand \BibitemShut  [1]{\csname bibitem#1\endcsname}%
\let\auto@bib@innerbib\@empty
%</preamble>
\bibitem [{\citenamefont {Kitagawa}\ \emph {et~al.}(2004)\citenamefont
  {Kitagawa}, \citenamefont {Takenaka},\ and\ \citenamefont
  {Fukuyama}}]{kitagawa2004applications}%
  \BibitemOpen
  \bibfield  {author} {\bibinfo {author} {\bibfnamefont {S.}~\bibnamefont
  {Kitagawa}}, \bibinfo {author} {\bibfnamefont {M.}~\bibnamefont {Takenaka}},
  \ and\ \bibinfo {author} {\bibfnamefont {Y.}~\bibnamefont {Fukuyama}},\
  }\href@noop {} {\bibfield  {journal} {\bibinfo  {journal} {Fuji Electric
  Review}\ }\textbf {\bibinfo {volume} {50}},\ \bibinfo {pages} {89} (\bibinfo
  {year} {2004})}\BibitemShut {NoStop}%
\bibitem [{\citenamefont {Odili}(2017)}]{odili2017applications}%
  \BibitemOpen
  \bibfield  {author} {\bibinfo {author} {\bibfnamefont {J.~B.}\ \bibnamefont
  {Odili}},\ }\href@noop {} {\bibfield  {journal} {\bibinfo  {journal} {Current
  Science}\ }\textbf {\bibinfo {volume} {113}},\ \bibinfo {pages} {2268}
  (\bibinfo {year} {2017})}\BibitemShut {NoStop}%
\bibitem [{\citenamefont {Paschos}(2014)}]{paschos2014applications}%
  \BibitemOpen
  \bibfield  {author} {\bibinfo {author} {\bibfnamefont {V.~T.}\ \bibnamefont
  {Paschos}},\ }\href@noop {} {\emph {\bibinfo {title} {Applications of
  Combinatorial Optimization}}}\ (\bibinfo  {publisher} {John Wiley \& Sons},\
  \bibinfo {year} {2014})\BibitemShut {NoStop}%
\bibitem [{\citenamefont {Barahona}(1982)}]{Barahona1982}%
  \BibitemOpen
  \bibfield  {author} {\bibinfo {author} {\bibfnamefont {F.}~\bibnamefont
  {Barahona}},\ }\href@noop {} {\bibfield  {journal} {\bibinfo  {journal}
  {Journal of Physics A: Mathematical and General}\ }\textbf {\bibinfo {volume}
  {15}},\ \bibinfo {pages} {3241} (\bibinfo {year} {1982})}\BibitemShut
  {NoStop}%
\bibitem [{\citenamefont {De~las Cuevas}\ and\ \citenamefont
  {Cubitt}(2016)}]{Cubitt_universality}%
  \BibitemOpen
  \bibfield  {author} {\bibinfo {author} {\bibfnamefont {G.}~\bibnamefont
  {De~las Cuevas}}\ and\ \bibinfo {author} {\bibfnamefont {T.~S.}\ \bibnamefont
  {Cubitt}},\ }\href@noop {} {\bibfield  {journal} {\bibinfo  {journal}
  {Science}\ }\textbf {\bibinfo {volume} {351}},\ \bibinfo {pages} {1180}
  (\bibinfo {year} {2016})}\BibitemShut {NoStop}%
\bibitem [{\citenamefont {Lucas}(2014)}]{lucas2014ising}%
  \BibitemOpen
  \bibfield  {author} {\bibinfo {author} {\bibfnamefont {A.}~\bibnamefont
  {Lucas}},\ }\href@noop {} {\bibfield  {journal} {\bibinfo  {journal}
  {Frontiers in physics}\ }\textbf {\bibinfo {volume} {2}},\ \bibinfo {pages}
  {5} (\bibinfo {year} {2014})}\BibitemShut {NoStop}%
\bibitem [{\citenamefont {Babaeian}\ \emph {et~al.}(2019)\citenamefont
  {Babaeian}, \citenamefont {Nguyen}, \citenamefont {Demir}, \citenamefont
  {Akbulut}, \citenamefont {Blanche}, \citenamefont {Kaneda}, \citenamefont
  {Guha}, \citenamefont {Neifeld},\ and\ \citenamefont
  {Peyghambarian}}]{babaeian2019single}%
  \BibitemOpen
  \bibfield  {author} {\bibinfo {author} {\bibfnamefont {M.}~\bibnamefont
  {Babaeian}}, \bibinfo {author} {\bibfnamefont {D.~T.}\ \bibnamefont
  {Nguyen}}, \bibinfo {author} {\bibfnamefont {V.}~\bibnamefont {Demir}},
  \bibinfo {author} {\bibfnamefont {M.}~\bibnamefont {Akbulut}}, \bibinfo
  {author} {\bibfnamefont {P.-A.}\ \bibnamefont {Blanche}}, \bibinfo {author}
  {\bibfnamefont {Y.}~\bibnamefont {Kaneda}}, \bibinfo {author} {\bibfnamefont
  {S.}~\bibnamefont {Guha}}, \bibinfo {author} {\bibfnamefont {M.~A.}\
  \bibnamefont {Neifeld}}, \ and\ \bibinfo {author} {\bibfnamefont
  {N.}~\bibnamefont {Peyghambarian}},\ }\href@noop {} {\bibfield  {journal}
  {\bibinfo  {journal} {Nature communications}\ }\textbf {\bibinfo {volume}
  {10}},\ \bibinfo {pages} {1} (\bibinfo {year} {2019})}\BibitemShut {NoStop}%
\bibitem [{\citenamefont {Pal}\ \emph {et~al.}(2020)\citenamefont {Pal},
  \citenamefont {Mahler}, \citenamefont {Tradonsky}, \citenamefont {Friesem},\
  and\ \citenamefont {Davidson}}]{pal2020rapid}%
  \BibitemOpen
  \bibfield  {author} {\bibinfo {author} {\bibfnamefont {V.}~\bibnamefont
  {Pal}}, \bibinfo {author} {\bibfnamefont {S.}~\bibnamefont {Mahler}},
  \bibinfo {author} {\bibfnamefont {C.}~\bibnamefont {Tradonsky}}, \bibinfo
  {author} {\bibfnamefont {A.~A.}\ \bibnamefont {Friesem}}, \ and\ \bibinfo
  {author} {\bibfnamefont {N.}~\bibnamefont {Davidson}},\ }\href@noop {}
  {\bibfield  {journal} {\bibinfo  {journal} {Physical Review Research}\
  }\textbf {\bibinfo {volume} {2}},\ \bibinfo {pages} {033008} (\bibinfo {year}
  {2020})}\BibitemShut {NoStop}%
\bibitem [{\citenamefont {Parto}\ \emph {et~al.}(2020)\citenamefont {Parto},
  \citenamefont {Hayenga}, \citenamefont {Marandi}, \citenamefont
  {Christodoulides},\ and\ \citenamefont {Khajavikhan}}]{parto2020realizing}%
  \BibitemOpen
  \bibfield  {author} {\bibinfo {author} {\bibfnamefont {M.}~\bibnamefont
  {Parto}}, \bibinfo {author} {\bibfnamefont {W.}~\bibnamefont {Hayenga}},
  \bibinfo {author} {\bibfnamefont {A.}~\bibnamefont {Marandi}}, \bibinfo
  {author} {\bibfnamefont {D.~N.}\ \bibnamefont {Christodoulides}}, \ and\
  \bibinfo {author} {\bibfnamefont {M.}~\bibnamefont {Khajavikhan}},\
  }\href@noop {} {\bibfield  {journal} {\bibinfo  {journal} {Nature materials}\
  }\textbf {\bibinfo {volume} {19}},\ \bibinfo {pages} {725} (\bibinfo {year}
  {2020})}\BibitemShut {NoStop}%
\bibitem [{\citenamefont {Yamamoto}\ \emph {et~al.}(2017)\citenamefont
  {Yamamoto}, \citenamefont {Aihara}, \citenamefont {Leleu}, \citenamefont
  {Kawarabayashi}, \citenamefont {Kako}, \citenamefont {Fejer}, \citenamefont
  {Inoue},\ and\ \citenamefont {Takesue}}]{yamamoto2017coherent}%
  \BibitemOpen
  \bibfield  {author} {\bibinfo {author} {\bibfnamefont {Y.}~\bibnamefont
  {Yamamoto}}, \bibinfo {author} {\bibfnamefont {K.}~\bibnamefont {Aihara}},
  \bibinfo {author} {\bibfnamefont {T.}~\bibnamefont {Leleu}}, \bibinfo
  {author} {\bibfnamefont {K.-i.}\ \bibnamefont {Kawarabayashi}}, \bibinfo
  {author} {\bibfnamefont {S.}~\bibnamefont {Kako}}, \bibinfo {author}
  {\bibfnamefont {M.}~\bibnamefont {Fejer}}, \bibinfo {author} {\bibfnamefont
  {K.}~\bibnamefont {Inoue}}, \ and\ \bibinfo {author} {\bibfnamefont
  {H.}~\bibnamefont {Takesue}},\ }\href@noop {} {\bibfield  {journal} {\bibinfo
   {journal} {npj Quantum Information}\ }\textbf {\bibinfo {volume} {3}},\
  \bibinfo {pages} {1} (\bibinfo {year} {2017})}\BibitemShut {NoStop}%
\bibitem [{\citenamefont {Inagaki}\ \emph {et~al.}(2016)\citenamefont
  {Inagaki}, \citenamefont {Haribara}, \citenamefont {Igarashi}, \citenamefont
  {Sonobe}, \citenamefont {Tamate}, \citenamefont {Honjo}, \citenamefont
  {Marandi}, \citenamefont {McMahon}, \citenamefont {Umeki}, \citenamefont
  {Enbutsu} \emph {et~al.}}]{inagaki2016coherent}%
  \BibitemOpen
  \bibfield  {author} {\bibinfo {author} {\bibfnamefont {T.}~\bibnamefont
  {Inagaki}}, \bibinfo {author} {\bibfnamefont {Y.}~\bibnamefont {Haribara}},
  \bibinfo {author} {\bibfnamefont {K.}~\bibnamefont {Igarashi}}, \bibinfo
  {author} {\bibfnamefont {T.}~\bibnamefont {Sonobe}}, \bibinfo {author}
  {\bibfnamefont {S.}~\bibnamefont {Tamate}}, \bibinfo {author} {\bibfnamefont
  {T.}~\bibnamefont {Honjo}}, \bibinfo {author} {\bibfnamefont
  {A.}~\bibnamefont {Marandi}}, \bibinfo {author} {\bibfnamefont {P.~L.}\
  \bibnamefont {McMahon}}, \bibinfo {author} {\bibfnamefont {T.}~\bibnamefont
  {Umeki}}, \bibinfo {author} {\bibfnamefont {K.}~\bibnamefont {Enbutsu}},
  \emph {et~al.},\ }\href@noop {} {\bibfield  {journal} {\bibinfo  {journal}
  {Science}\ }\textbf {\bibinfo {volume} {354}},\ \bibinfo {pages} {603}
  (\bibinfo {year} {2016})}\BibitemShut {NoStop}%
\bibitem [{\citenamefont {McMahon}\ \emph {et~al.}(2016)\citenamefont
  {McMahon}, \citenamefont {Marandi}, \citenamefont {Haribara}, \citenamefont
  {Hamerly}, \citenamefont {Langrock}, \citenamefont {Tamate}, \citenamefont
  {Inagaki}, \citenamefont {Takesue}, \citenamefont {Utsunomiya}, \citenamefont
  {Aihara} \emph {et~al.}}]{mcmahon2016fully}%
  \BibitemOpen
  \bibfield  {author} {\bibinfo {author} {\bibfnamefont {P.~L.}\ \bibnamefont
  {McMahon}}, \bibinfo {author} {\bibfnamefont {A.}~\bibnamefont {Marandi}},
  \bibinfo {author} {\bibfnamefont {Y.}~\bibnamefont {Haribara}}, \bibinfo
  {author} {\bibfnamefont {R.}~\bibnamefont {Hamerly}}, \bibinfo {author}
  {\bibfnamefont {C.}~\bibnamefont {Langrock}}, \bibinfo {author}
  {\bibfnamefont {S.}~\bibnamefont {Tamate}}, \bibinfo {author} {\bibfnamefont
  {T.}~\bibnamefont {Inagaki}}, \bibinfo {author} {\bibfnamefont
  {H.}~\bibnamefont {Takesue}}, \bibinfo {author} {\bibfnamefont
  {S.}~\bibnamefont {Utsunomiya}}, \bibinfo {author} {\bibfnamefont
  {K.}~\bibnamefont {Aihara}},  \emph {et~al.},\ }\href@noop {} {\bibfield
  {journal} {\bibinfo  {journal} {Science}\ }\textbf {\bibinfo {volume}
  {354}},\ \bibinfo {pages} {614} (\bibinfo {year} {2016})}\BibitemShut
  {NoStop}%
\bibitem [{\citenamefont {Johnson}\ \emph {et~al.}(2011)\citenamefont
  {Johnson}, \citenamefont {Amin}, \citenamefont {Gildert}, \citenamefont
  {Lanting}, \citenamefont {Hamze}, \citenamefont {Dickson}, \citenamefont
  {Harris}, \citenamefont {Berkley}, \citenamefont {Johansson}, \citenamefont
  {Bunyk} \emph {et~al.}}]{johnson2011quantum}%
  \BibitemOpen
  \bibfield  {author} {\bibinfo {author} {\bibfnamefont {M.~W.}\ \bibnamefont
  {Johnson}}, \bibinfo {author} {\bibfnamefont {M.~H.}\ \bibnamefont {Amin}},
  \bibinfo {author} {\bibfnamefont {S.}~\bibnamefont {Gildert}}, \bibinfo
  {author} {\bibfnamefont {T.}~\bibnamefont {Lanting}}, \bibinfo {author}
  {\bibfnamefont {F.}~\bibnamefont {Hamze}}, \bibinfo {author} {\bibfnamefont
  {N.}~\bibnamefont {Dickson}}, \bibinfo {author} {\bibfnamefont
  {R.}~\bibnamefont {Harris}}, \bibinfo {author} {\bibfnamefont {A.~J.}\
  \bibnamefont {Berkley}}, \bibinfo {author} {\bibfnamefont {J.}~\bibnamefont
  {Johansson}}, \bibinfo {author} {\bibfnamefont {P.}~\bibnamefont {Bunyk}},
  \emph {et~al.},\ }\href@noop {} {\bibfield  {journal} {\bibinfo  {journal}
  {Nature}\ }\textbf {\bibinfo {volume} {473}},\ \bibinfo {pages} {194}
  (\bibinfo {year} {2011})}\BibitemShut {NoStop}%
\bibitem [{\citenamefont {Denchev}\ \emph {et~al.}(2016)\citenamefont
  {Denchev}, \citenamefont {Boixo}, \citenamefont {Isakov}, \citenamefont
  {Ding}, \citenamefont {Babbush}, \citenamefont {Smelyanskiy}, \citenamefont
  {Martinis},\ and\ \citenamefont {Neven}}]{denchev2016computational}%
  \BibitemOpen
  \bibfield  {author} {\bibinfo {author} {\bibfnamefont {V.~S.}\ \bibnamefont
  {Denchev}}, \bibinfo {author} {\bibfnamefont {S.}~\bibnamefont {Boixo}},
  \bibinfo {author} {\bibfnamefont {S.~V.}\ \bibnamefont {Isakov}}, \bibinfo
  {author} {\bibfnamefont {N.}~\bibnamefont {Ding}}, \bibinfo {author}
  {\bibfnamefont {R.}~\bibnamefont {Babbush}}, \bibinfo {author} {\bibfnamefont
  {V.}~\bibnamefont {Smelyanskiy}}, \bibinfo {author} {\bibfnamefont
  {J.}~\bibnamefont {Martinis}}, \ and\ \bibinfo {author} {\bibfnamefont
  {H.}~\bibnamefont {Neven}},\ }\href@noop {} {\bibfield  {journal} {\bibinfo
  {journal} {Physical Review X}\ }\textbf {\bibinfo {volume} {6}},\ \bibinfo
  {pages} {031015} (\bibinfo {year} {2016})}\BibitemShut {NoStop}%
\bibitem [{\citenamefont {Arute}\ \emph {et~al.}(2019)\citenamefont {Arute},
  \citenamefont {Arya}, \citenamefont {Babbush}, \citenamefont {Bacon},
  \citenamefont {Bardin}, \citenamefont {Barends}, \citenamefont {Biswas},
  \citenamefont {Boixo}, \citenamefont {Brandao}, \citenamefont {Buell} \emph
  {et~al.}}]{arute2019quantum}%
  \BibitemOpen
  \bibfield  {author} {\bibinfo {author} {\bibfnamefont {F.}~\bibnamefont
  {Arute}}, \bibinfo {author} {\bibfnamefont {K.}~\bibnamefont {Arya}},
  \bibinfo {author} {\bibfnamefont {R.}~\bibnamefont {Babbush}}, \bibinfo
  {author} {\bibfnamefont {D.}~\bibnamefont {Bacon}}, \bibinfo {author}
  {\bibfnamefont {J.~C.}\ \bibnamefont {Bardin}}, \bibinfo {author}
  {\bibfnamefont {R.}~\bibnamefont {Barends}}, \bibinfo {author} {\bibfnamefont
  {R.}~\bibnamefont {Biswas}}, \bibinfo {author} {\bibfnamefont
  {S.}~\bibnamefont {Boixo}}, \bibinfo {author} {\bibfnamefont {F.~G.}\
  \bibnamefont {Brandao}}, \bibinfo {author} {\bibfnamefont {D.~A.}\
  \bibnamefont {Buell}},  \emph {et~al.},\ }\href@noop {} {\bibfield  {journal}
  {\bibinfo  {journal} {Nature}\ }\textbf {\bibinfo {volume} {574}},\ \bibinfo
  {pages} {505} (\bibinfo {year} {2019})}\BibitemShut {NoStop}%
\bibitem [{\citenamefont {Cai}\ \emph {et~al.}(2020)\citenamefont {Cai},
  \citenamefont {Kumar}, \citenamefont {Van~Vaerenbergh}, \citenamefont
  {Sheng}, \citenamefont {Liu}, \citenamefont {Li}, \citenamefont {Liu},
  \citenamefont {Foltin}, \citenamefont {Yu}, \citenamefont {Xia} \emph
  {et~al.}}]{cai2020power}%
  \BibitemOpen
  \bibfield  {author} {\bibinfo {author} {\bibfnamefont {F.}~\bibnamefont
  {Cai}}, \bibinfo {author} {\bibfnamefont {S.}~\bibnamefont {Kumar}}, \bibinfo
  {author} {\bibfnamefont {T.}~\bibnamefont {Van~Vaerenbergh}}, \bibinfo
  {author} {\bibfnamefont {X.}~\bibnamefont {Sheng}}, \bibinfo {author}
  {\bibfnamefont {R.}~\bibnamefont {Liu}}, \bibinfo {author} {\bibfnamefont
  {C.}~\bibnamefont {Li}}, \bibinfo {author} {\bibfnamefont {Z.}~\bibnamefont
  {Liu}}, \bibinfo {author} {\bibfnamefont {M.}~\bibnamefont {Foltin}},
  \bibinfo {author} {\bibfnamefont {S.}~\bibnamefont {Yu}}, \bibinfo {author}
  {\bibfnamefont {Q.}~\bibnamefont {Xia}},  \emph {et~al.},\ }\href@noop {}
  {\bibfield  {journal} {\bibinfo  {journal} {Nature Electronics}\ }\textbf
  {\bibinfo {volume} {3}},\ \bibinfo {pages} {409} (\bibinfo {year}
  {2020})}\BibitemShut {NoStop}%
\bibitem [{\citenamefont {Kim}\ \emph {et~al.}(2010)\citenamefont {Kim},
  \citenamefont {Chang}, \citenamefont {Korenblit}, \citenamefont {Islam},
  \citenamefont {Edwards}, \citenamefont {Freericks}, \citenamefont {Lin},
  \citenamefont {Duan},\ and\ \citenamefont {Monroe}}]{kim2010quantum}%
  \BibitemOpen
  \bibfield  {author} {\bibinfo {author} {\bibfnamefont {K.}~\bibnamefont
  {Kim}}, \bibinfo {author} {\bibfnamefont {M.-S.}\ \bibnamefont {Chang}},
  \bibinfo {author} {\bibfnamefont {S.}~\bibnamefont {Korenblit}}, \bibinfo
  {author} {\bibfnamefont {R.}~\bibnamefont {Islam}}, \bibinfo {author}
  {\bibfnamefont {E.~E.}\ \bibnamefont {Edwards}}, \bibinfo {author}
  {\bibfnamefont {J.~K.}\ \bibnamefont {Freericks}}, \bibinfo {author}
  {\bibfnamefont {G.-D.}\ \bibnamefont {Lin}}, \bibinfo {author} {\bibfnamefont
  {L.-M.}\ \bibnamefont {Duan}}, \ and\ \bibinfo {author} {\bibfnamefont
  {C.}~\bibnamefont {Monroe}},\ }\href@noop {} {\bibfield  {journal} {\bibinfo
  {journal} {Nature}\ }\textbf {\bibinfo {volume} {465}},\ \bibinfo {pages}
  {590} (\bibinfo {year} {2010})}\BibitemShut {NoStop}%
\bibitem [{\citenamefont {Berloff}\ \emph {et~al.}(2017)\citenamefont
  {Berloff}, \citenamefont {Silva}, \citenamefont {Kalinin}, \citenamefont
  {Askitopoulos}, \citenamefont {T{\"o}pfer}, \citenamefont {Cilibrizzi},
  \citenamefont {Langbein},\ and\ \citenamefont
  {Lagoudakis}}]{berloff2017realizing}%
  \BibitemOpen
  \bibfield  {author} {\bibinfo {author} {\bibfnamefont {N.~G.}\ \bibnamefont
  {Berloff}}, \bibinfo {author} {\bibfnamefont {M.}~\bibnamefont {Silva}},
  \bibinfo {author} {\bibfnamefont {K.}~\bibnamefont {Kalinin}}, \bibinfo
  {author} {\bibfnamefont {A.}~\bibnamefont {Askitopoulos}}, \bibinfo {author}
  {\bibfnamefont {J.~D.}\ \bibnamefont {T{\"o}pfer}}, \bibinfo {author}
  {\bibfnamefont {P.}~\bibnamefont {Cilibrizzi}}, \bibinfo {author}
  {\bibfnamefont {W.}~\bibnamefont {Langbein}}, \ and\ \bibinfo {author}
  {\bibfnamefont {P.~G.}\ \bibnamefont {Lagoudakis}},\ }\href@noop {}
  {\bibfield  {journal} {\bibinfo  {journal} {Nature materials}\ } (\bibinfo
  {year} {2017})}\BibitemShut {NoStop}%
\bibitem [{\citenamefont {Kalinin}\ \emph {et~al.}(2020)\citenamefont
  {Kalinin}, \citenamefont {Amo}, \citenamefont {Bloch},\ and\ \citenamefont
  {Berloff}}]{kalinin2020polaritonic}%
  \BibitemOpen
  \bibfield  {author} {\bibinfo {author} {\bibfnamefont {K.~P.}\ \bibnamefont
  {Kalinin}}, \bibinfo {author} {\bibfnamefont {A.}~\bibnamefont {Amo}},
  \bibinfo {author} {\bibfnamefont {J.}~\bibnamefont {Bloch}}, \ and\ \bibinfo
  {author} {\bibfnamefont {N.~G.}\ \bibnamefont {Berloff}},\ }\href@noop {}
  {\bibfield  {journal} {\bibinfo  {journal} {Nanophotonics}\ }\textbf
  {\bibinfo {volume} {9}},\ \bibinfo {pages} {4127} (\bibinfo {year}
  {2020})}\BibitemShut {NoStop}%
\bibitem [{\citenamefont {Vretenar}\ \emph {et~al.}(2021)\citenamefont
  {Vretenar}, \citenamefont {Kassenberg}, \citenamefont {Bissesar},
  \citenamefont {Toebes},\ and\ \citenamefont
  {Klaers}}]{vretenar2021controllable}%
  \BibitemOpen
  \bibfield  {author} {\bibinfo {author} {\bibfnamefont {M.}~\bibnamefont
  {Vretenar}}, \bibinfo {author} {\bibfnamefont {B.}~\bibnamefont
  {Kassenberg}}, \bibinfo {author} {\bibfnamefont {S.}~\bibnamefont
  {Bissesar}}, \bibinfo {author} {\bibfnamefont {C.}~\bibnamefont {Toebes}}, \
  and\ \bibinfo {author} {\bibfnamefont {J.}~\bibnamefont {Klaers}},\
  }\href@noop {} {\bibfield  {journal} {\bibinfo  {journal} {Physical Review
  Research}\ }\textbf {\bibinfo {volume} {3}},\ \bibinfo {pages} {023167}
  (\bibinfo {year} {2021})}\BibitemShut {NoStop}%
\bibitem [{\citenamefont {Vadlamani}\ \emph {et~al.}(2020)\citenamefont
  {Vadlamani}, \citenamefont {Xiao},\ and\ \citenamefont
  {Yablonovitch}}]{vadlamani2020physics}%
  \BibitemOpen
  \bibfield  {author} {\bibinfo {author} {\bibfnamefont {S.~K.}\ \bibnamefont
  {Vadlamani}}, \bibinfo {author} {\bibfnamefont {T.~P.}\ \bibnamefont {Xiao}},
  \ and\ \bibinfo {author} {\bibfnamefont {E.}~\bibnamefont {Yablonovitch}},\
  }\href@noop {} {\bibfield  {journal} {\bibinfo  {journal} {Proceedings of the
  National Academy of Sciences}\ }\textbf {\bibinfo {volume} {117}},\ \bibinfo
  {pages} {26639} (\bibinfo {year} {2020})}\BibitemShut {NoStop}%
\bibitem [{\citenamefont {Hauke}\ \emph {et~al.}(2020)\citenamefont {Hauke},
  \citenamefont {Katzgraber}, \citenamefont {Lechner}, \citenamefont
  {Nishimori},\ and\ \citenamefont {Oliver}}]{hauke2020perspectives}%
  \BibitemOpen
  \bibfield  {author} {\bibinfo {author} {\bibfnamefont {P.}~\bibnamefont
  {Hauke}}, \bibinfo {author} {\bibfnamefont {H.~G.}\ \bibnamefont
  {Katzgraber}}, \bibinfo {author} {\bibfnamefont {W.}~\bibnamefont {Lechner}},
  \bibinfo {author} {\bibfnamefont {H.}~\bibnamefont {Nishimori}}, \ and\
  \bibinfo {author} {\bibfnamefont {W.~D.}\ \bibnamefont {Oliver}},\
  }\href@noop {} {\bibfield  {journal} {\bibinfo  {journal} {Reports on
  Progress in Physics}\ }\textbf {\bibinfo {volume} {83}},\ \bibinfo {pages}
  {054401} (\bibinfo {year} {2020})}\BibitemShut {NoStop}%
\bibitem [{\citenamefont {Kamaletdinov}\ and\ \citenamefont
  {Berloff}(2021)}]{kamaletdinov2021quantized}%
  \BibitemOpen
  \bibfield  {author} {\bibinfo {author} {\bibfnamefont {A.}~\bibnamefont
  {Kamaletdinov}}\ and\ \bibinfo {author} {\bibfnamefont {N.~G.}\ \bibnamefont
  {Berloff}},\ }\href@noop {} {\bibfield  {journal} {\bibinfo  {journal} {arXiv
  preprint arXiv:2109.05867}\ } (\bibinfo {year} {2021})}\BibitemShut {NoStop}%
\bibitem [{\citenamefont {Goto}(2016)}]{goto2016bifurcation}%
  \BibitemOpen
  \bibfield  {author} {\bibinfo {author} {\bibfnamefont {H.}~\bibnamefont
  {Goto}},\ }\href@noop {} {\bibfield  {journal} {\bibinfo  {journal}
  {Scientific reports}\ }\textbf {\bibinfo {volume} {6}},\ \bibinfo {pages} {1}
  (\bibinfo {year} {2016})}\BibitemShut {NoStop}%
\bibitem [{\citenamefont {Tatsumura}\ \emph {et~al.}(2019)\citenamefont
  {Tatsumura}, \citenamefont {Dixon},\ and\ \citenamefont
  {Goto}}]{tatsumura2019fpga}%
  \BibitemOpen
  \bibfield  {author} {\bibinfo {author} {\bibfnamefont {K.}~\bibnamefont
  {Tatsumura}}, \bibinfo {author} {\bibfnamefont {A.~R.}\ \bibnamefont
  {Dixon}}, \ and\ \bibinfo {author} {\bibfnamefont {H.}~\bibnamefont {Goto}},\
  }in\ \href@noop {} {\emph {\bibinfo {booktitle} {2019 29th International
  Conference on Field Programmable Logic and Applications (FPL)}}}\ (\bibinfo
  {organization} {IEEE},\ \bibinfo {year} {2019})\ pp.\ \bibinfo {pages}
  {59--66}\BibitemShut {NoStop}%
\bibitem [{\citenamefont {Goto}\ \emph {et~al.}(2021)\citenamefont {Goto},
  \citenamefont {Endo}, \citenamefont {Suzuki}, \citenamefont {Sakai},
  \citenamefont {Kanao}, \citenamefont {Hamakawa}, \citenamefont {Hidaka},
  \citenamefont {Yamasaki},\ and\ \citenamefont {Tatsumura}}]{goto2021high}%
  \BibitemOpen
  \bibfield  {author} {\bibinfo {author} {\bibfnamefont {H.}~\bibnamefont
  {Goto}}, \bibinfo {author} {\bibfnamefont {K.}~\bibnamefont {Endo}}, \bibinfo
  {author} {\bibfnamefont {M.}~\bibnamefont {Suzuki}}, \bibinfo {author}
  {\bibfnamefont {Y.}~\bibnamefont {Sakai}}, \bibinfo {author} {\bibfnamefont
  {T.}~\bibnamefont {Kanao}}, \bibinfo {author} {\bibfnamefont
  {Y.}~\bibnamefont {Hamakawa}}, \bibinfo {author} {\bibfnamefont
  {R.}~\bibnamefont {Hidaka}}, \bibinfo {author} {\bibfnamefont
  {M.}~\bibnamefont {Yamasaki}}, \ and\ \bibinfo {author} {\bibfnamefont
  {K.}~\bibnamefont {Tatsumura}},\ }\href@noop {} {\bibfield  {journal}
  {\bibinfo  {journal} {Science Advances}\ }\textbf {\bibinfo {volume} {7}},\
  \bibinfo {pages} {eabe7953} (\bibinfo {year} {2021})}\BibitemShut {NoStop}%
\bibitem [{\citenamefont {Kalinin}\ and\ \citenamefont
  {Berloff}(2019)}]{kalinin2019polaritonic}%
  \BibitemOpen
  \bibfield  {author} {\bibinfo {author} {\bibfnamefont {K.~P.}\ \bibnamefont
  {Kalinin}}\ and\ \bibinfo {author} {\bibfnamefont {N.~G.}\ \bibnamefont
  {Berloff}},\ }\href@noop {} {\bibfield  {journal} {\bibinfo  {journal}
  {Physical Review B}\ }\textbf {\bibinfo {volume} {100}},\ \bibinfo {pages}
  {245306} (\bibinfo {year} {2019})}\BibitemShut {NoStop}%
\bibitem [{\citenamefont {Wang}\ \emph {et~al.}(2013)\citenamefont {Wang},
  \citenamefont {Marandi}, \citenamefont {Wen}, \citenamefont {Byer},\ and\
  \citenamefont {Yamamoto}}]{wang2013coherent}%
  \BibitemOpen
  \bibfield  {author} {\bibinfo {author} {\bibfnamefont {Z.}~\bibnamefont
  {Wang}}, \bibinfo {author} {\bibfnamefont {A.}~\bibnamefont {Marandi}},
  \bibinfo {author} {\bibfnamefont {K.}~\bibnamefont {Wen}}, \bibinfo {author}
  {\bibfnamefont {R.~L.}\ \bibnamefont {Byer}}, \ and\ \bibinfo {author}
  {\bibfnamefont {Y.}~\bibnamefont {Yamamoto}},\ }\href@noop {} {\bibfield
  {journal} {\bibinfo  {journal} {Physical Review A}\ }\textbf {\bibinfo
  {volume} {88}},\ \bibinfo {pages} {063853} (\bibinfo {year}
  {2013})}\BibitemShut {NoStop}%
\bibitem [{\citenamefont {Leleu}\ \emph {et~al.}(2020)\citenamefont {Leleu},
  \citenamefont {Khoyratee}, \citenamefont {Levi}, \citenamefont {Hamerly},
  \citenamefont {Kohno},\ and\ \citenamefont {Aihara}}]{leleu2020chaotic}%
  \BibitemOpen
  \bibfield  {author} {\bibinfo {author} {\bibfnamefont {T.}~\bibnamefont
  {Leleu}}, \bibinfo {author} {\bibfnamefont {F.}~\bibnamefont {Khoyratee}},
  \bibinfo {author} {\bibfnamefont {T.}~\bibnamefont {Levi}}, \bibinfo {author}
  {\bibfnamefont {R.}~\bibnamefont {Hamerly}}, \bibinfo {author} {\bibfnamefont
  {T.}~\bibnamefont {Kohno}}, \ and\ \bibinfo {author} {\bibfnamefont
  {K.}~\bibnamefont {Aihara}},\ }\href@noop {} {\bibfield  {journal} {\bibinfo
  {journal} {arXiv e-prints}\ ,\ \bibinfo {pages} {arXiv}} (\bibinfo {year}
  {2020})}\BibitemShut {NoStop}%
\bibitem [{\citenamefont {Stroev}\ and\ \citenamefont
  {Berloff}(2021)}]{stroev2021discrete}%
  \BibitemOpen
  \bibfield  {author} {\bibinfo {author} {\bibfnamefont {N.}~\bibnamefont
  {Stroev}}\ and\ \bibinfo {author} {\bibfnamefont {N.~G.}\ \bibnamefont
  {Berloff}},\ }\href@noop {} {\bibfield  {journal} {\bibinfo  {journal}
  {Physical Review Letters}\ }\textbf {\bibinfo {volume} {126}},\ \bibinfo
  {pages} {050504} (\bibinfo {year} {2021})}\BibitemShut {NoStop}%
\bibitem [{\citenamefont {Hoppensteadt}\ and\ \citenamefont
  {Izhikevich}(1996{\natexlab{a}})}]{hoppensteadt1996synaptic}%
  \BibitemOpen
  \bibfield  {author} {\bibinfo {author} {\bibfnamefont {F.~C.}\ \bibnamefont
  {Hoppensteadt}}\ and\ \bibinfo {author} {\bibfnamefont {E.~M.}\ \bibnamefont
  {Izhikevich}},\ }\href@noop {} {\bibfield  {journal} {\bibinfo  {journal}
  {Biological cybernetics}\ }\textbf {\bibinfo {volume} {75}},\ \bibinfo
  {pages} {117} (\bibinfo {year} {1996}{\natexlab{a}})}\BibitemShut {NoStop}%
\bibitem [{\citenamefont {Hoppensteadt}\ and\ \citenamefont
  {Izhikevich}(1996{\natexlab{b}})}]{hoppensteadt1996synapticB}%
  \BibitemOpen
  \bibfield  {author} {\bibinfo {author} {\bibfnamefont {F.~C.}\ \bibnamefont
  {Hoppensteadt}}\ and\ \bibinfo {author} {\bibfnamefont {E.~M.}\ \bibnamefont
  {Izhikevich}},\ }\href@noop {} {\bibfield  {journal} {\bibinfo  {journal}
  {Biological Cybernetics}\ }\textbf {\bibinfo {volume} {75}},\ \bibinfo
  {pages} {129} (\bibinfo {year} {1996}{\natexlab{b}})}\BibitemShut {NoStop}%
\bibitem [{\citenamefont {Hoppensteadt}\ and\ \citenamefont
  {Izhikevich}(1997)}]{hoppensteadt1997weakly}%
  \BibitemOpen
  \bibfield  {author} {\bibinfo {author} {\bibfnamefont {F.~C.}\ \bibnamefont
  {Hoppensteadt}}\ and\ \bibinfo {author} {\bibfnamefont {E.~M.}\ \bibnamefont
  {Izhikevich}},\ }\href@noop {} {\emph {\bibinfo {title} {Weakly connected
  neural networks}}},\ Vol.\ \bibinfo {volume} {126}\ (\bibinfo  {publisher}
  {Springer Science \& Business Media},\ \bibinfo {year} {1997})\BibitemShut
  {NoStop}%
\bibitem [{\citenamefont {Hoppensteadt}\ and\ \citenamefont
  {Izhikevich}(2001)}]{hoppensteadt2001synchronization}%
  \BibitemOpen
  \bibfield  {author} {\bibinfo {author} {\bibfnamefont {F.~C.}\ \bibnamefont
  {Hoppensteadt}}\ and\ \bibinfo {author} {\bibfnamefont {E.~M.}\ \bibnamefont
  {Izhikevich}},\ }\href@noop {} {\bibfield  {journal} {\bibinfo  {journal}
  {IEEE Transactions on Circuits and Systems I: Fundamental Theory and
  Applications}\ }\textbf {\bibinfo {volume} {48}},\ \bibinfo {pages} {133}
  (\bibinfo {year} {2001})}\BibitemShut {NoStop}%
\bibitem [{\citenamefont {Nesterov}(2003)}]{nesterov2003introductory}%
  \BibitemOpen
  \bibfield  {author} {\bibinfo {author} {\bibfnamefont {Y.}~\bibnamefont
  {Nesterov}},\ }\href@noop {} {\emph {\bibinfo {title} {Introductory lectures
  on convex optimization: A basic course}}},\ Vol.~\bibinfo {volume} {87}\
  (\bibinfo  {publisher} {Springer Science \& Business Media},\ \bibinfo {year}
  {2003})\BibitemShut {NoStop}%
\bibitem [{\citenamefont {Su}\ \emph {et~al.}(2014)\citenamefont {Su},
  \citenamefont {Boyd},\ and\ \citenamefont {Candes}}]{su2014differential}%
  \BibitemOpen
  \bibfield  {author} {\bibinfo {author} {\bibfnamefont {W.}~\bibnamefont
  {Su}}, \bibinfo {author} {\bibfnamefont {S.}~\bibnamefont {Boyd}}, \ and\
  \bibinfo {author} {\bibfnamefont {E.}~\bibnamefont {Candes}},\ }\href@noop {}
  {\bibfield  {journal} {\bibinfo  {journal} {Advances in neural information
  processing systems}\ }\textbf {\bibinfo {volume} {27}} (\bibinfo {year}
  {2014})}\BibitemShut {NoStop}%
\bibitem [{\citenamefont {Celledoni}\ \emph {et~al.}(2021)\citenamefont
  {Celledoni}, \citenamefont {Ehrhardt}, \citenamefont {Etmann}, \citenamefont
  {McLachlan}, \citenamefont {Owren}, \citenamefont {SCHONLIEB},\ and\
  \citenamefont {Sherry}}]{celledoni2021structure}%
  \BibitemOpen
  \bibfield  {author} {\bibinfo {author} {\bibfnamefont {E.}~\bibnamefont
  {Celledoni}}, \bibinfo {author} {\bibfnamefont {M.~J.}\ \bibnamefont
  {Ehrhardt}}, \bibinfo {author} {\bibfnamefont {C.}~\bibnamefont {Etmann}},
  \bibinfo {author} {\bibfnamefont {R.~I.}\ \bibnamefont {McLachlan}}, \bibinfo
  {author} {\bibfnamefont {B.}~\bibnamefont {Owren}}, \bibinfo {author}
  {\bibfnamefont {C.-B.}\ \bibnamefont {SCHONLIEB}}, \ and\ \bibinfo {author}
  {\bibfnamefont {F.}~\bibnamefont {Sherry}},\ }\href@noop {} {\bibfield
  {journal} {\bibinfo  {journal} {European Journal of Applied Mathematics}\
  }\textbf {\bibinfo {volume} {32}},\ \bibinfo {pages} {888} (\bibinfo {year}
  {2021})}\BibitemShut {NoStop}%
\bibitem [{\citenamefont {Kalinin}\ and\ \citenamefont
  {Berloff}(2018{\natexlab{a}})}]{kalinin2018gain}%
  \BibitemOpen
  \bibfield  {author} {\bibinfo {author} {\bibfnamefont {K.~P.}\ \bibnamefont
  {Kalinin}}\ and\ \bibinfo {author} {\bibfnamefont {N.~G.}\ \bibnamefont
  {Berloff}},\ }\href@noop {} {\bibfield  {journal} {\bibinfo  {journal} {arXiv
  preprint arXiv:1805.01371}\ } (\bibinfo {year}
  {2018}{\natexlab{a}})}\BibitemShut {NoStop}%
\bibitem [{\citenamefont {Kalinin}\ and\ \citenamefont
  {Berloff}(2018{\natexlab{b}})}]{kalinin2018networks}%
  \BibitemOpen
  \bibfield  {author} {\bibinfo {author} {\bibfnamefont {K.~P.}\ \bibnamefont
  {Kalinin}}\ and\ \bibinfo {author} {\bibfnamefont {N.~G.}\ \bibnamefont
  {Berloff}},\ }\href@noop {} {\bibfield  {journal} {\bibinfo  {journal} {New
  Journal of Physics}\ }\textbf {\bibinfo {volume} {20}},\ \bibinfo {pages}
  {113023} (\bibinfo {year} {2018}{\natexlab{b}})}\BibitemShut {NoStop}%
\bibitem [{\citenamefont {Kalinin}\ and\ \citenamefont
  {Berloff}(2018{\natexlab{c}})}]{kalinin2018global}%
  \BibitemOpen
  \bibfield  {author} {\bibinfo {author} {\bibfnamefont {K.~P.}\ \bibnamefont
  {Kalinin}}\ and\ \bibinfo {author} {\bibfnamefont {N.~G.}\ \bibnamefont
  {Berloff}},\ }\href@noop {} {\bibfield  {journal} {\bibinfo  {journal}
  {Scientific reports}\ }\textbf {\bibinfo {volume} {8}},\ \bibinfo {pages}
  {17791} (\bibinfo {year} {2018}{\natexlab{c}})}\BibitemShut {NoStop}%
\bibitem [{\citenamefont {Izhikevich}(2000)}]{izhikevich2000computing}%
  \BibitemOpen
  \bibfield  {author} {\bibinfo {author} {\bibfnamefont {E.~M.}\ \bibnamefont
  {Izhikevich}},\ }\href@noop {} {\bibfield  {journal} {\bibinfo  {journal}
  {Neural Networks}\ }\textbf {\bibinfo {volume} {5255}},\ \bibinfo {pages} {1}
  (\bibinfo {year} {2000})}\BibitemShut {NoStop}%
\end{thebibliography}%

%-----------------

%\end{thebibliography}

\appendix*

\end{document}